\documentstyle[12pt,fleqn,epsfig]{article}
\begin{document}
      
\begin{center}
{\bf INSTITUT~F\"{U}R~KERNPHYSIK,~UNIVERSIT\"{A}T~FRANKFURT}\\
D - 60486 Frankfurt, August--Euler--Strasse 6, Germany
\end{center}

\hfill IKF--HENPG/02--00


\vspace{1.5cm}
\begin{center}
{\Large \bf 
Charm Estimate
from the Dilepton Spectra 
}
\end{center}  
\begin{center}
{\Large \bf 
in Nuclear Collisions
}
\end{center}  

\vspace{1cm}

\begin{center}

Marek
Ga\'zdzicki$^{a,}$\footnote{E--mail: marek@ikf.physik.uni--frankfurt.de}
and
Mark I. Gorenstein$^{b,}$\footnote{E--mail: goren@ap3.bitp.kiev.ua}
\end{center}

\vspace{0.5cm}

$^a$ CERN, Geneva, Switzerland,   

~~~Institut f\"ur Kernphysik, 
Universit\"at Frankfurt, Frankfurt, Germany\\

$^b$ Bogolyubov Institute for Theoretical Physics, Kiev, Ukraine,

~~~Institut f\"ur  
Theoretische Physik, Universit\"at Frankfurt,
Germany

\vspace{1cm}

\begin{abstract}
\noindent
A validity of a recent estimate of an upper limit of charm production
in central Pb+Pb collisions at 158 A$\cdot$GeV 
is critically
discussed.
Within a simple model
we study
properties of the background subtraction procedure  used
for an extraction  of the charm signal from the analysis
of dilepton spectra.
We demonstrate that a  production asymmetry between
positively and negatively charged 
background muons and a large multiplicity
of signal pairs leads to biased results.
Therefore the applicability of this procedure for the analysis
of nucleus--nucleus data should be reconsidered before final
conclusions on the upper limit estimate of charm production 
could be drawn.
\end{abstract}

\newpage
Measurement of the invariant mass spectra of opposite--sign
lepton pairs (dileptons) allow to extract information
otherwise difficult or even impossible to obtain.
Among interesting processes which contribute to dilepton production
are decays of vector mesons ($\rho,~w,~\phi,~J/\psi,~\psi^{'} $),
Drell--Yan as well as thermal creation of dileptons, and decays of charm
hadrons.
Decays of pions and kaons are a dominant source of 
uninteresting (background) dileptons which should be subtracted before
deconvolution
of contributions from the interesting (signal) sources is performed.

Recent analysis of dimuon spectrum measured in central Pb+Pb collisions
at 158 A$\cdot$GeV by NA50 Collaboration \cite{NA50} 
suggests a significant enhancement of dilepton
production in the intermidiate mass region (1.5$\div$2.5 GeV) over
the standard sources. 
The primary interpretation attributes this observation to the
increased production of open charm \cite{NA50}.
In the following theoretical papers other possible sources of the
observed effect are proposed which do not invoke enhancement of
the open charm yield \cite{TH}.
This suggests to interpret the NA50 result as an estimate of the
upper limit (about 3 times above pQCD predictions) 
of open charm multiplicity in Pb+Pb collisions at SPS.
The above conclusion relies, however, on the assumption that the background
subtraction procedure used to extract signal sources  
gives unbiased results.
In this work we show that this  assumption is questionable.
In particular, an  asymmetry
in the production of positively and negatively charged background
dileptons
and a high multiplicity of signal pairs lead to
the result which differs from the one usually assumed
in the data interpretation.
Our analysis is done within a simple model  
based on the assumptions
used to justify
the background subtraction procedure \cite{NA50}.

\vspace{0.3cm}

In  central Pb+Pb collisions at SPS, due to high
multiplicity of produced hadrons,
the multiplicity of background dileptons is much higher 
($\approx$ 95 \%)
than the multiplicity of signal pairs ($\approx$ 5 \%).
The invariant mass spectra of the Drell--Yan, thermal, and open charm
contributions are broad and essentially structureless.
Consequently their extraction requires very precise knowledge
of the shape and the absolute normalisation of the background
distribution.
The necessary accuracy can not be reached by 
calculation of the background based on a model.
Therefore
in order to decrease the systematic error
of the background estimation a method   based on
the measured data was developed  
and used in the analysis of  dilepton spectra \cite{NA50, CERES, HELIOS}.
In this method
the background contribution to  dilepton spectra
is calculated  as 
$2 \sqrt{ \langle n_{++} \rangle \langle n_{--} \rangle }$,
where $\langle n_{++} \rangle$ and $\langle n_{--} \rangle$ are measured
multiplicities of like--sign lepton pairs.

The NA50 experiment measured the mean multiplicity 
of like--sign, $\langle n_{++} \rangle$ and $\langle n_{--} \rangle$, 
and opposite--sign,  $\langle n_{+-} \rangle$, muon pairs.
One usually distinguishes two classes of muons:
the "independent" muons coming from decays of pions and kaons
($h$ mesons)
and the "correlated" muons originating from vector meson decays,
Drell--Yan and thermal creation of dimuons, and from decays of pairs of charm
hadrons. 
For simplicity of the initial considerations let us assume that
the correlated muons only come from  the decays of charm hadrons, which
we denote here by $D$ and 
$\overline{D}$.
The  meaning in which the words
"independent" and "correlated" used above  is the following.
Let $N_+$, $N_-$ be the numbers 
of positively and negatively charged hadrons 
(kaons and/or pions)  produced 
in a given nucleus--nucleus (A+A) collision.
The numbers $N_+$, $N_-$ are  {\it independent} when the probability 
to observe them can be factorized:
\begin{equation}\label{prod}
P(N_+,N_-)~=~P_+(N_+) \times P_-(N_-)~,
\end{equation}
where  $P_+(N_+)$ and $P_-(N_-)$ are the probability distributions for
independent observation of $N_+$ or $N_-$ hadrons.
Due to charm conservation the numbers of 
$D$ and $\overline{D}$ hadrons are expected to be equal
in each event ($N_D$ =
$N_{\overline{D}}$);  
the production of $D$ and $\overline{D}$ hadrons 
is {\it correlated}.
The independence or the correlation of muon sources
leads to an independence or a correlation of muons
originating from these sources. 
The assumption of approximately {\it independent} $K^+$ and $K^-$ 
(or $\pi^+$ and $\pi^-$) production in  A+A event
is justified by large number of different hadron species created
in the collision. 
Then, e.g. the electric charge and
strangeness of produced $K^+$ in a
given event  could be in fact compensated by many
different hadron combinations, not just only by $K^-$.

Let us denote by
$\alpha_h$ and $\alpha_D$  the probabilities 
that a decay
of a single $h$ or $D$   leads to a muon 
inside the NA50 spectrometer.
In an  event with multiplicities  $N_+$, $N_-$ and $N_D$ 
the probabilities 
to observe $n$ muons of a given sort
are binominaly distributed:
\begin{equation}\label{p1}
P_i(n_+^i)~=~\frac{N_+!}{n_+^i!~(N_+ - n_+^i)!}~(\alpha_h)^{n_+^i}~
(1 - \alpha_h)^{N_+ ~-~ n_+^i}~,
\end{equation}
\begin{equation}\label{p2}     
P_i(n_-^i)~=~\frac{N_-!}{n_-^i!~(N_- - n_-^i)!}~(\alpha_h)^{n_-^i}~
(1 - \alpha_h)^{N_- ~-~ n_-^i}~.
\end{equation}
\begin{equation}\label{p3}     
P_c(n_+^c)~=~\frac{N_D!}{n_+^c!~(N_D - n_+^c)!}~(\alpha_D)^{n_+^c}~
(1 - \alpha_D)^{N_D~ -~ n_+^c}~,
\end{equation}
\begin{equation}\label{p4}
P_c(n_-^c)~=~\frac{N_D!}{n_-^c!~(N_D - n_-^c)!}~(\alpha_D)^{n_-^c}~
(1 - \alpha_D)^{N_D~ -~ n_-^c}~.
\end{equation}
where  $n_+^i$, $n_-^i$, $n_+^c$ and $n_-^c$ are numbers of positively
and negatively charged muons from "independent" 
and "correlated" sources. From 
Eqs.~(\ref{p1}-\ref{p4}) one finds
\begin{eqnarray}\label{av2bin}\label{avbin}
\overline{n_+^i}~&=&~\alpha_h~N_+~,~~~~
\overline{n_-^i}~=~\alpha_h~N_-~,~~~~ 
\overline{n_+^c}~=~\overline{n_-^c}~=~\alpha_D~N_D~,\\
\overline{(n_+^i)^2}~&=&~\alpha_h~(1-\alpha_h)~N_+~+~\alpha_h^2~N_+^2~,\\
\overline{(n_-^i)^2}~&=&~\alpha_h~(1-\alpha_h)~N_-~+~\alpha_h^2~N_-^2~,\\
\overline{(n_+^c)^2}~&=&~
\overline{(n_-^c)^2}~=~ \alpha_D~(1-\alpha_D)~N_D~+~\alpha_D^2~N_D^2~.
\end{eqnarray}

\noindent
We introduce now the probabilities, $A_h$, $A_D$, 
and $A_{hD}$
that  muon pairs from, respectively, $hh$, $DD$ and $hD$ decays 
are detected 
within the
{\it dimuon} acceptance.
These probabilities depend on cuts on the dimuon properties and,
for given experimental cuts, on momentum spectra
of dimuon sources.
Assuming that the probabilities $A$ are multiplicity independent,
we arrive at the following expressions for the numbers 
of like--sign and opposite--sign muon pairs, for 
{\it fixed values} of $N_+$, $N_-$ and $N_D$ 
\begin{eqnarray}\label{++}
\overline{n_{++}}~&=&~
A_h~\sum_{n_+^i}~\frac{n_+^i(n_+^i - 1)}{2}~P_i(n_+^i)~
+~A_D~\sum_{n_+^c}~\frac{n_+^c(n_+^c - 1)}{2}~P_c(n_+^c)~ \\
&+&~A_{hD}~\sum_{n_+^i,n_+^c}~n_+^i~n_+^c ~P_i(n_+^i)~P_c(n_+^c)
\nonumber \\
&=&~\frac{A_h}{2}~\left(\overline{(n_+^i)^2} - \overline{n_+^i}\right) 
~+~\frac{A_D}{2}~\left(\overline{(n_+^c)^2} - \overline{n_+^c}\right)~+~
A_{hD}~\overline{n_+^i}~\overline{n_+^c} \nonumber\\
&=&~\frac{A_h}{2}~\alpha_h^2~\left(N_+^2 - N_+\right)~+~
\frac{A_D}{2}~\alpha_D^2~\left(N_D^2 - N_D\right)
~+~A_{hD}~\alpha_h\alpha_D~N_+~ N_D~,\nonumber
\end{eqnarray}
\begin{equation}\label{--}
\overline{n_{--}}~=~
\frac{A_h}{2}~\alpha_h^2~\left(N_-^2 - N_-\right)~+~
\frac{A_D}{2}~\alpha_D^2~\left(N_D^2 - N_D\right)
~+~A_{hD}~\alpha_h\alpha_D~N_-~ N_D~,
\end{equation}
\begin{eqnarray}\label{+-}
\overline{n_{+-}}~&=&~
A_h~\sum_{n_+^i,n_-^i}~n_+^i n_-^i ~P_i(n_+^i) ~P_i(n_-^i)~
+~A_D~\sum_{n_+^c,n_-^c} n_+^c n_-^c ~P_c(n_+^c)~ P_c(n_-^c)\\
&+&~A_{hD}~\sum_{n_+^i,n_-^c}~n_+^i~n_-^c ~P_i(n_+^i)~P_c(n_-^c)~+~
A_{hD}~\sum_{n_-^i,n_+^c}~n_-^i~n_+^c ~P_i(n_-^i)~P_c(n_+^c)
\nonumber \\
&=&~A_h~\alpha_h^2~N_+~ N_- ~+~
A_D~\alpha_D^2~N_D^2  
~+~A_{hD}~\alpha_h \alpha_D~N_D~(N_+ + N_-)~.\nonumber
\end{eqnarray}
Here we have made a simplified assumption that the shape of 
momentum spectra of $h^+$ and $h^-$ (as well as $D$
and $\overline{D}$) are similar and, therefore,
$A_h^{++}=A_h^{--}=A_h^{+-}\equiv A_h$, 
$A_{hD}^{++}=A_{hD}^{--}=A_{hD}^{+-}\equiv A_{hD}$ and
$A_D^{++}=A_D^{--}=A_D^{+-}\equiv A_D$
(the last equation means that possible momentum correlations between
$D$ and $\overline{D}$ are also neglected).
Note that if there are no cuts on the dimuon properties the above
probabilities become equal to unity,
$A_h=A_{hD}=A_D=1$, i.e. assuming all $A$-probabilities equal
to one in Eqs.~(\ref{++}-\ref{+-}) we count all possible
dimuon pairs. However, as soon as one fixes some dimuon properties 
(e.g. an invariant mass of the dimuon pair) all $A$-probabilities
are evidently smaller than unity and their actual numerical values
become dependent on the shape of $h$ and $D$ momentum spectra and
their decay kinematics.
Note also that in Eqs.~(\ref{++}-\ref{+-}) an
{\it independence} of muon numbers $n_+^i$ and $n_-^i$ is
due to assumed in Eq.~(\ref{prod}) independence of $N_+$ and $N_-$
which entered into $P_i(n_+^i)$ (\ref{p1})
and $P_i(n_-^i)$ (\ref{p2}).
A {\it correlation} of muon numbers $n_+^c$ and $n_-^c$ is
due to the correlation of $N_D$ and $N_{\overline{D}}$
($N_D=N_{\overline{D}}$) which entered into $P_c(n_+^c)$
(\ref{p3}) and $P_c(n_-^c)$ (\ref{p4}) probability distributions.
The correlation of $n_+^c$ and $n_-^c$ is of course weaker than that
for $N_D$ and $N_{\overline{D}}$, so that $n_+^c$ are not necessarily
equal to $n_-^c$ in each event. 
 
In order to find the final mean multiplicities of the dimuons one should
average the obtained numbers over all possible values of $N_+,N_-,N_D$.
To simplify the following calculations we assume that
the relevant multiplicity  distributions are Poisson distributions
\begin{equation}\label{poiss}
P(N)~=~\frac{\overline{N}^N}{N!}~\exp(- \overline{N})~.
\end{equation}
In this case one gets:
\begin{eqnarray}\label{++av}
\langle n_{++} \rangle~ & = &~ \sum_{N_+,N_-,N_D}~
\overline{n_{++}}~~P(N_+)~P(N_-)~P(N_D) ~=~
\frac{1}{2}~A_h~\alpha_h^2~\left(\overline{N_+}\right)^2 \\
&+&~
\frac{1}{2}~A_D~\alpha_D^2~\left(\overline{N_D}\right)^2~+~
A_{hD}~\alpha_h~\alpha_D~\overline{N_+}~~\overline{N_D}~.\nonumber
\end{eqnarray}
\begin{eqnarray}\label{--av}
\langle n_{--} \rangle~ & = &~ \sum_{N_+,N_-,N_D}~
\overline{n_{--}}~~P(N_+)~P(N_-)~P(N_D) ~=~
\frac{1}{2}~A_h~\alpha_h^2~\left(\overline{N_-}\right)^2 \\
&+&~
\frac{1}{2}~A_D~\alpha_D^2~\left(\overline{N_D}\right)^2~+~
A_{hD}~\alpha_h \alpha_D~\overline{N_-}~~\overline{N_D}~.\nonumber
\end{eqnarray}
\begin{eqnarray}\label{+-av}
\langle n_{+-} \rangle~ & = &~ \sum_{N_+,N_-,N_D}~
\overline{n_{+-}}~~P(N_+)~P(N_-)~P(N_D)~=~ 
A_h~\alpha_h^2~\overline{N_+}~~\overline{N_-} \\
&+&~
A_D~\alpha_D^2~\left[\left(\overline{N_D}\right)^2~+~\overline{N_D}\right]~+~
A_{hD}~\alpha_h~\alpha_D~\overline{N_D}~\left(\overline{N_+}~+~
\overline{N_-}\right)~.
\nonumber
\end{eqnarray}
Note again that $N_D=N_{\overline{D}}$ is assumed in each event
and, therefore, there is no independent summation over $N_{\overline{D}}$
in the above equations.
Eqs.~(\ref{++av}-\ref{+-av}) can be rewritten as
\begin{eqnarray}
\langle n_{++} \rangle~ & = &~ 
\frac{1}{2}~a_h~h_+^2~+~
\frac{1}{2}~a_d~D^2~+~a_m~h_+~D~,\label{++f}\\
\langle n_{--} \rangle~ & = &~ 
\frac{1}{2}~a_h~h_-^2~+~ 
\frac{1}{2}~a_d~D^2~+~a_m~h_-~D~, \label{--f}\\
\langle n_{+-} \rangle~ & = &~ 
a_h~h_+~h_-~+~a_d~D^2~+~a_d~D~+ 
~a_m~D~\left(h_+ ~+~ h_- \right)~,\label{+-f}
\end{eqnarray}
by introducing the following notations:
\begin{eqnarray}\label{notation}
a_h ~& \equiv & ~A_h~\alpha_h^2~,~~~
a_d ~\equiv ~A_D~\alpha_D^2~,~~~
a_m ~\equiv ~A_{hD}~\alpha_h~\alpha_D ~,\\
\overline{N_+}~&\equiv& ~h_+~,~~~
\overline{N_-}~\equiv ~h_-~,~~~
\overline{N_D}~\equiv ~D~.
\end{eqnarray}
Parameters $a_h$, $a_d$ and $a_m$ are therefore
the probabilities to observe two muons from the corresponding hadron
sources (these probabilities are $\alpha_h^2$, $\alpha_D^2$ and
$\alpha_h\alpha_D$) within experimental cuts on muon pair properties
(these cuts lead to additional factors $A_h$, $A_D$ and $A_{hD}$).

\vspace{0.5cm}
In the experimental procedure the {\it background} contribution to the
dimuon spectrum is calculated as:
\begin{equation}\label{bgr}
\langle n_{+-}^{Bgr} \rangle~  \equiv ~
2~\sqrt{\langle n_{++} \rangle~ \langle n_{--} \rangle}~.
\end{equation}
The number of signal $(\mu^+,\mu^-)$--pairs 
is assumed to be: 
\begin{equation}\label{sgl}
\langle n_{+-}^{Sgl} \rangle~ \equiv ~
\langle n_{+-} \rangle~-~
\langle n_{+-}^{Bgr} \rangle~.
\end{equation}
It is expected that the subtraction procedure (\ref{sgl})
cancels out all false $(\mu^+,\mu^-)$--pairs i.e. the pairs from $hh$ and
$hD$ decays, and that 
$\langle n_{+-}^{Sgl} \rangle$
is  proportional to the multiplicity  of $D$ hadrons:
\begin{equation}\label{sglD}
\langle n_{+-}^{Sgl} \rangle~=  ~a_d~D~.
\end{equation}

\vspace{0.5cm}   
Let us consider some properties of the subtraction procedure (\ref{sgl})
by discussing two simple examples within the 
model.

\vspace{0.3cm}
\noindent
{\bf Example 1:}
We assume that there is no contribution from $D$-decays. 
In our model
this assumption can be introduced  by setting
$\alpha_D=0$.
Consequently  $a_d=a_m=0$
and Eqs.~(\ref{++f}-\ref{+-f}) result in:
\begin{equation}\label{D0}   
\langle n_{++} \rangle~=~\frac{1}{2}~a_h~h_+^2~,~~~
\langle n_{+-} \rangle~=~\frac{1}{2}~a_h~h_-^2~,~~~
\langle n_{+-} \rangle~=~a_h~h_+~h_-~.
\end{equation}
Using Eq.~(\ref{sgl}) one obtains that  $\langle n_{+-}^{Sgl} \rangle =0$,
i.e. the measured signal multiplicity is equal to zero as expected in the
case of absence of dimuons from the correlated source. 
This result is valid for any value of
$h_+$ and $h_-$. 

\vspace{0.3cm}
\noindent
{\bf Example 2:} 
In this example we assume that there are correlated dimuons
$a_d~D > 0$
but the number of positively and negatively charged 
background hadrons is equal
($h_+ = h_- \equiv h$).
Under these conditions  Eqs.~(\ref{++f}-\ref{+-f}) can be rewritten as
\begin{eqnarray}
\langle n_{++} \rangle~ & = &~\langle n_{--} \rangle~=~
\frac{1}{2}~a_h~h^2~+~  
\frac{1}{2}~a_d~D^2~+~a_m~h~D~,\label{++0}\\
\langle n_{+-} \rangle~ & = &~
a_h~h^2~+~a_d~D^2~+~a_d~D~+
~2~a_m~h~D~.\label{+-0}
\end{eqnarray}
Eq.~(\ref{sgl}) gives 
$\langle n_{+-}^{Sgl} \rangle =  a_d~D$
which agrees exactly with the expectation (\ref{sglD}).

\vspace{0.3cm}
Finally we consider
the general case, i.e. $a_d~D > 0$ and 
$h_+ \neq h_-$. This last condition corresponds to
the relation between pion and kaon average multiplicities
measured 
in heavy ion collisions: $\langle \pi^- \rangle > \langle
\pi^+ \rangle$ and $\langle K^+ \rangle > \langle K^- \rangle$. From 
Eqs.~(\ref{++f}-\ref{sgl}) 
by straightforward calculations one finds
\begin{equation}\label{sgl1}
\langle n_{+-}^{Sgl} \rangle~=~\langle n_{+-} \rangle~-~
\sqrt{\left(\langle n_{+-} \rangle ~-~ a_d~D\right)^2~+~\gamma~D^2}~,
\end{equation}
where
$$
\gamma ~\equiv~
\left(a_h~a_d~-~a_m^2\right)~\left(h_+~-~h_-\right)^2~.
$$

\vspace{0.3cm}
\noindent
It is easy to see that for
$\alpha_D=0$  
and/or $h_+=h_-=h$ one gets $\gamma =0$,
and the results obtained in Examples 1 and 2 are reproduced.
We repeat that in the absence of cuts
on the dimuon properties one has $A_h=A_{hD}=A_D=1$. Therefore,
$a_h~a_d~-~a_m^2=0$ (i.e. $\gamma =0$) and consequently we have again
unbiased estimate of the mean multiplicity of D mesons.  
In  general, however, the result  differs from the
expected one  (\ref{sglD}).
The presence of experimental cuts on dimuons
(e.g. one fixes the dimuon invariant mass
in the region $M=1.5\div 2.5$~GeV) causes that the probabilities
$A_h,A_{hD}$ and $A_D$ are smaller than unity,
destroys the equality $A_hA_D=A_{hD}^2$ and,
therefore, leads to non-zero value of $\gamma$.
With cuts on dimuon properties the experimental number of signal pairs 
$\langle n_{+-}^{Sgl}\rangle$ is not equal to $a_d D$.
By fitting  $a_d D^*$ to $\langle n_{+-}^{Sgl}\rangle$ 
one finds the spurious number of $D$ hadrons which we denoted by $D^*$.
There are two distinct cases.

\vspace{0.3cm}
\noindent
{\bf Case 1:} ~~~ $a_h~a_d~-~a_m^2 < 0~, ~~(\gamma < 0)~.$ \\
The experimentally 
measured signal, $\langle n_{+-}^{Sgl}\rangle$ (\ref{sgl1}),
is larger than the expected value $a_d D$ and therefore 
the extracted spurious number of $D$ hadrons is larger
than the true one
($D<D^*$).

\vspace{0.3cm}
\noindent
{\bf Case 2:} ~~~ $a_h~a_d~-~a_m^2 > 0~, ~~ (\gamma > 0)~.$ \\
The experimentally
measured signal, $\langle n_{+-}^{Sgl}\rangle$ (\ref{sgl1}),
is smaller than the expected value $a_d D$ and therefore
the extracted spurious number of $D$ hadrons is smaller
than the true one
($D>D^*$).

\vspace{0.5cm}
In the NA50 analysis of the dimuon spectra in terms of the open
charm enhancement the used background subtraction 
procedure was checked
for two different cases.
First of all, it was shown to work correctly for simulated
central Pb+Pb collisions at 158 A$\cdot$GeV.
However in this simulation correlated (signal) 
muon sources were not included.
Thus this check is equivalent to our Example 1, for which the procedure
works exactly.
Secondly the open charm yield was extracted for p+A interactions
and it was shown to agree with the yield from  direct measurements.
Eq.~(\ref{sgl1}) and  Example 1 show that the deviation from the
expected result decreases with  decreasing multiplicity of
$D$ hadrons.
Thus  the success of the procedure applied to p+A interactions does
not proof its applicability to Pb+Pb collisions in which multiplicity
of $D$ hadrons may be higher
even by a factor of about $10^4$ \cite{Go:98,Ma:99}.

Note that our results are obtained in a highly simplified
model. The assumptions concerning independent production of
background muons (Eq.~(\ref{prod})),
the Poissonian multiplicity
distributions of
hadrons (Eq.~(\ref{poiss}))
and the absence of $D$ meson momentum correlations
($A_D^{++}=A_D^{--}=A_D^{+-}\equiv A_D$)
seem to be
questionable or even incorrect.
Discussion of the possible {\it additional biases} introduced
by these effects is beyond the scope of this paper.
We also do not attempt here to calculate numerical values of
$A_h,A_{hD},A_D$ for the specific NA50 experimental acceptance
of dimuon pairs. 

We close the paper by concluding that the applicability of the
background subtraction procedure widely used in the analysis
of dilepton spectra in nucleus--nucleus collisions
should be reconsidered.
In particular final statement on the upper limit
of the open charm multiplicity in central Pb+Pb
collisions at 158 A$\cdot$GeV resulting from the 
analysis of the dimuon spectrum requiers
further studies 
in order to quantify
a  magnitude of the bias.
They should include numerical simulations of the specific
experimental set--up and consider various  particle
production models.

\vspace{1cm}
\noindent
{\bf Acknowledgements}

We thank M. Botje, K.A Bugaev, D. Jouan, M. van Leeuwen, 
E. Scomparin,  P. Seyboth
and P. Sonderegger for discussions and comments to the manuscript.
We acknowledge the
financial support of BMBF and  DFG, Germany.
The research described in this publication was made possible in part by
Award No. UP1-2119 of the U.S. Civilian Research \& Development
Foundation for the Independent States of the Former Soviet Union
(CRDF).


\end{document}